\documentclass[11pt,a4paper]{article}
\usepackage[T1]{fontenc}
\usepackage{lmodern}
\usepackage[utf8]{inputenc}
\usepackage[margin=25mm]{geometry}
\usepackage{graphicx,xurl,booktabs,longtable,array,float,placeins}
\usepackage[authoryear,round]{natbib}
\usepackage{microtype}
\usepackage[hidelinks]{hyperref}
\usepackage{orcidlink}
\hypersetup{pdftitle={Training for sustainable scientific software in the AI era: from code production to scientific stewardship},pdfauthor={Juliette Alimena}}
\newcommand{\doi}[1]{\unskip\space\href{https://doi.org/#1}{\nolinkurl{doi:#1}}}

\title{Training for sustainable scientific software in the AI era: from code production to scientific stewardship}
\author{Juliette Alimena\,\orcidlink{0000-0001-6030-3191}\\[4pt]
\small Deutsches Elektronen-Synchrotron DESY, Hamburg, Germany\\
\small Correspondence: \href{mailto:juliette.alimena@desy.de}{juliette.alimena@desy.de}}
\date{}
\begin{document}
\maketitle
\begin{abstract}
Generative and agentic artificial intelligence (AI) systems are changing how researchers interact with scientific software. Their effects on productivity vary across tasks and users, but they can weaken the connection between producing a successful computational artifact and retaining scientific command over the consequential choices and evidence on which it depends. For high-energy physics, this shifts part of the training challenge from code production toward scientific stewardship: researchers must be able to validate, explain, reproduce, maintain, and justify software and workflows partly generated or modified by AI. We argue that AI-aware software training should therefore combine dedicated instruction on AI systems with the integration of AI into established software and analysis training. Testing, provenance, reproducibility, resource judgment, and scientific validation should be exercised in AI-assisted workflows, with competence judged by the evidence supporting a result rather than by code generation alone. We frame this as a problem of sustainable scientific software: AI-assisted workflows should remain trustworthy and maintainable, use computational resources responsibly, and avoid making core competence unnecessarily dependent on privileged access to particular tools or infrastructure.
\end{abstract}
\noindent\textbf{Keywords:} scientific software, software training, high-energy physics, generative AI, scientific stewardship, reproducibility, software sustainability

\section{Introduction}\label{sec:1}

High-energy physics (HEP) depends on researchers being able to work effectively with complex scientific software, but acquiring that competence has long required more than learning a programming language. Training programs within experiments, institutions, and the broader HEP software community therefore combine software fundamentals with domain-specific workflows, computing infrastructure, and analysis practice. Compact Muon Solenoid (CMS) Data Analysis Schools (CMSDAS) provide a collaboration-specific model: a week-long school in which about 80\% of school time is devoted to hands-on work that progresses from tools and physics objects to a 2.5-day intensive analysis exercise using real CMS data in teams of about 6–8 with close facilitator support \citep{r1,r21}. DESY’s sustainable scientific-computing workshops provide a complementary institution-level model across research communities \citep{r2}. The October 2026 two-day \emph{Computing Starter Kit and Green Coding} workshop at DESY and University of Hamburg (UHH) brought together about 40 in-person participants from particle physics and photon science for short instruction and hands-on exercises on Git, testing and continuous integration, containers, batch computing, storage, software maintainability, and efficient resource use, and added institutional artificial intelligence (AI) services and coding agents to this broader software-training context. These settings illustrate two complementary questions now facing HEP training: what researchers need to know specifically about AI systems, and how established scientific-software competencies should be exercised when AI becomes part of ordinary computational work.

Generative and increasingly agentic AI systems can assist with planning, coding, debugging, tool use, analysis, and documentation. As discussed in Section~\ref{sec:2}, their effects on productivity are heterogeneous across tasks and users. More consequential for scientific training, as we argue, is that successful production of a computational artifact can become less tightly coupled to the researcher’s command over the consequential choices, assumptions, and evidence on which it depends. Recent studies of AI-assisted programming report improved task completion alongside weaker understanding or verification of the resulting code. Educational reviews also emphasize scaffolding: structured guidance that is reduced as learners become more capable of working independently. As agents take on larger parts of computational workflows, this separation can extend beyond individual code fragments.

We argue that this changes the emphasis required in software training. Researchers must still learn to construct computational solutions, but they must increasingly also learn to take scientific responsibility for software and workflows they did not fully construct themselves. We use \textbf{scientific stewardship} to describe that responsibility: understanding consequential choices, validating outputs, preserving provenance and reproducibility, maintaining and reusing software appropriately, and exercising judgment about computational resources and tool use. These concerns are not unique to HEP, and related ideas are already emerging in the broader research-software community. This Perspective translates these ideas of scientific stewardship into concrete implications for HEP software training.

Here, \textbf{training for sustainable scientific software} means equipping researchers to steward workflows that remain trustworthy and maintainable, use computational resources responsibly, including consideration of environmental impact, and avoid unnecessary dependence on privileged commercial tools or infrastructure. We focus on training rather than comprehensive AI energy accounting, model development, ethics, or security, except where agentic and security issues sharpen the training problem. Drawing on experience from institutional sustainable-computing workshops and collaboration-specific CMS training, we examine what changes with AI (Section~\ref{sec:2}), what stewardship requires (Section~\ref{sec:3}), how training should adapt (Section~\ref{sec:4}), and the complementary roles of institutions, collaborations, and community efforts (Section~\ref{sec:5}).

\section{What changes when AI enters scientific software work?}\label{sec:2}

AI is changing software production, but its effects on productivity are neither uniform nor fully characterized. In three randomized field experiments involving 4,867 professional developers, access to an AI coding assistant increased completed tasks by about 26\% on average, with larger gains among less-experienced developers \citep{r3}. By contrast, a randomized study of experienced open-source developers working in mature repositories with which they were already familiar found that early-2025 AI tools made participants about 19\% slower \citep{r4}. A 2026 follow-up suggested that newer AI tools may now accelerate similar work, but substantial selection effects and difficulties measuring agentic workflows prevented a reliable quantitative estimate of the change \citep{r5}. Neither setting is necessarily representative of HEP. Together, however, the studies show that productivity effects depend strongly on task, user, software environment, and tool, so training should not assume universal gains.

Productivity measured as the time required to complete a predefined task may also miss a different potential effect. If AI lowers implementation effort, speculative or technically ambitious ideas may become feasible to test that would otherwise be abandoned as too costly. This could expand scientific exploration without necessarily shortening projects overall, while also increasing the volume and complexity of work requiring review and validation.

A more consequential issue for training is the relationship between successful task completion and scientific command over the resulting computation. In a 2026 crossover experiment with 69 students, researchers, and professional developers, median task completeness more than doubled with AI assistance, while median correctness on technical questions about the produced code fell by 12.5 percentage points, from 100\% without AI to 87.5\% with AI \citep{r6}. In undergraduate programming courses, students also performed significantly worse when asked to identify and correct errors in code generated by large language models than on instructor-designed programming tasks \citep{r7}. A systematic review of 72 empirical studies in computing education identifies a recurring completion-comprehension trade-off, while also finding that AI can support explanation, debugging, and problem solving when its use is appropriately structured \citep{r8}. The relevant distinction is therefore not simply between AI-assisted and unaided programming, but between obtaining a satisfactory artifact and retaining command over the assumptions, failure modes, and evidence needed to justify it.

Agentic systems extend this separation beyond code generation. Whereas an assistant may propose code or explanations, an agent can inspect a repository, invoke tools, execute commands, modify files, and iterate on its own outputs. Recent work on scientific agents emphasizes evaluation, documentation, reproducibility, hallucination, tool misuse, and human oversight as central concerns \citep{r12}. As delegation increases, researchers need to be able to establish what the system did, what information and tools it used, which consequential choices were made, and what evidence supports the resulting workflow.

This is not a transition from human-written software to fully automated science, nor does it make code production irrelevant. Instead, a working artifact becomes weaker evidence of command over the underlying computation, while human effort may shift toward architecture, planning, defining successful outcomes, and validation. If implementation capacity grows faster than review capacity, validation and scientific judgment become limiting resources. Scientific-software competence must therefore include responsibility for artifacts and workflows partly constructed by AI as well as the ability to construct solutions.

\section{What does scientific stewardship require?}\label{sec:3}

The competencies required for scientific stewardship are not new: research software already needs to be understandable, tested, reproducible, maintainable, and appropriate for its computational resources. AI changes the context in which these requirements must be met: researchers may increasingly take responsibility for workflows they did not construct in detail themselves. Similar concerns about trust, verification, provenance, maintenance, and long-term stewardship are emerging across the research-software community \citep{r9,r10}. We focus on three dimensions relevant to training: trustworthy and durable workflows, resource responsibility, and sustainable access and tool independence.

\textbf{Trustworthy and durable workflows.} AI-assisted workflows must remain understandable, testable, reproducible, and open to scrutiny. Researchers need not have written every line, but they should be able to identify consequential choices, judge whether the implementation fits the scientific problem, and establish what evidence supports it. Testing, documentation, version control, reproducible environments, and maintainability were already central to sustainable-computing training \citep{r2}; AI increases their importance. Agentic systems add a further requirement for provenance. Here, provenance means a record of how a computational artifact or result was produced that is sufficiently complete to reconstruct and audit consequential parts of the workflow. For consequential agent-assisted work, this can include the model or service and version where available, material prompts or version-controlled instructions, tool or action logs, commits or artifact identifiers, software dependencies, and validation steps. Full transcripts are not inherently required and may be inappropriate if they contain credentials, unpublished results, or information restricted by institutional or collaboration policy. Work on autonomous science likewise emphasizes auditable provenance, reproducible documentation, and evaluation \citep{r11,r12}.

\textbf{Resource responsibility.} A scientifically correct workflow can still waste resources through unnecessary computation, poor scaling, excessive data movement or storage, or inappropriate hardware. Sustainable practice therefore starts before optimization: avoid unnecessary work, test before large-scale production, reuse validated software or data products, and measure resource use before scaling. Sharing reusable data products, including generated event samples, can reduce duplicated computation and resource consumption \citep{r13}. AI does not have a uniformly positive or negative effect on this problem. It can help optimize software or replace costly calculations, but generating and operating AI models also consumes resources, and improvements in runtime do not necessarily imply equivalent reductions in energy use. For example, a 2025 study of code optimization using large language models found cases with substantial runtime and energy savings \citep{r14}. However, the energy cost of generating the optimization meant that energetic break-even could require hundreds to hundreds of thousands of subsequent executions. Faster implementations also did not consistently yield proportionate energy savings. The relevant training principle is therefore to measure the full workflow, including AI assistance where its resource use can be estimated, and choose models, deployment modes, and conventional computational methods proportionate to the scientific task rather than assuming any approach is intrinsically more sustainable.

\textbf{Sustainable access and tool independence.} Finally, effective AI use should not depend unnecessarily on privileged access to a particular commercial service, subscription tier, computing facility, or collaboration-specific infrastructure. Access to generative AI already varies across educational environments: a 2026 study of more than 20,000 university students in 58 countries found that technical access and the local institutional environment were important predictors of ChatGPT use \citep{r15}. This argues for separating durable scientific competencies from capabilities tied to a particular service. Institutionally provided systems can reduce dependence on individual subscriptions but create their own infrastructure dependencies; locally hosted or open-weight models provide another option where feasible, while shifting infrastructure and maintenance burdens to institutions. Collaboration or institutional rules may also restrict sending code, data, or unpublished results to external services. Exercises that work across multiple supported tools can make training more portable. Sustainable stewardship therefore concerns not only whether an AI-assisted workflow works today, but whether the knowledge, practices, and infrastructure needed to use such systems responsibly can be maintained and shared over time.

\begin{figure}[tbp]
\centering\includegraphics[width=\textwidth]{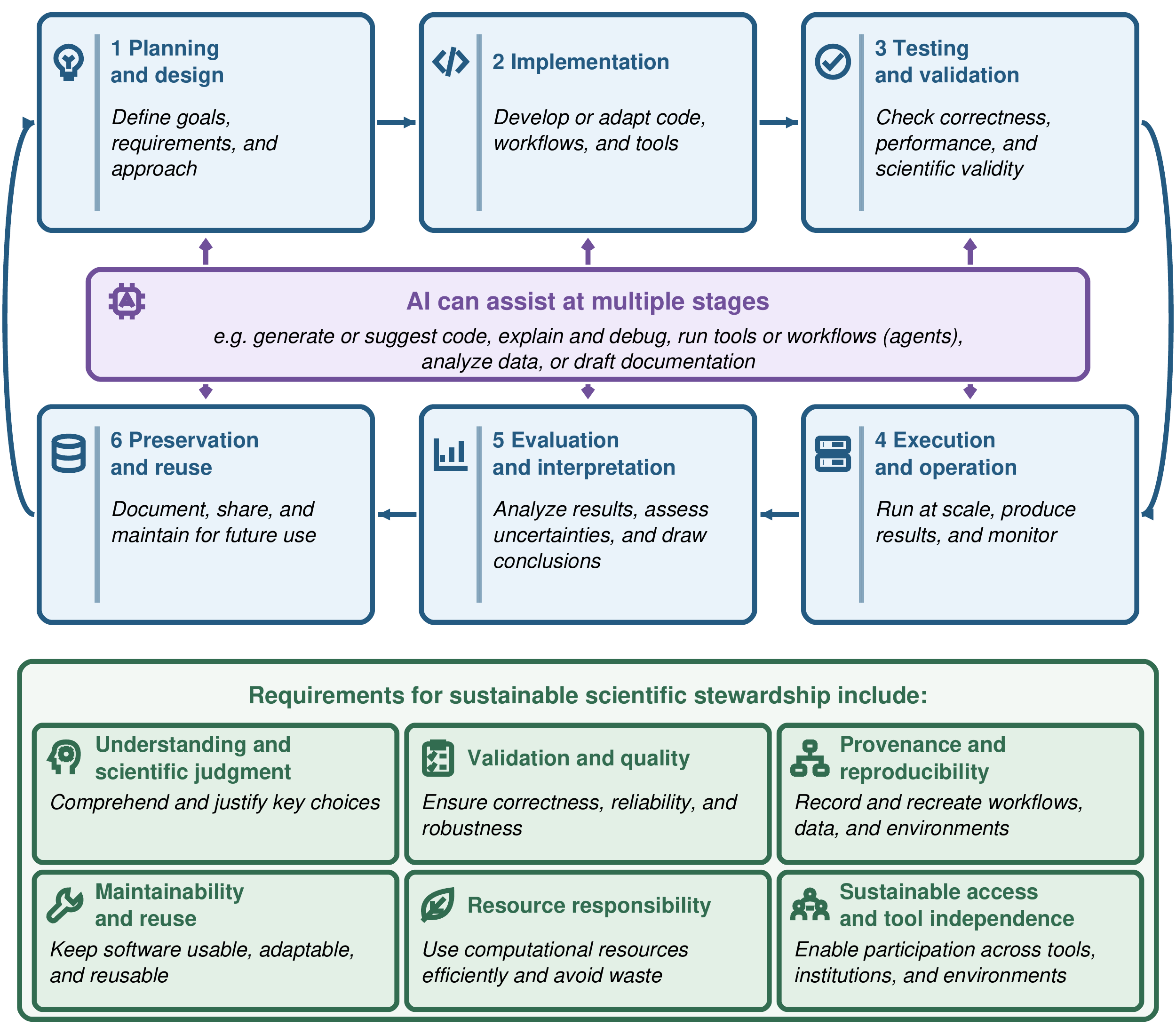}
\caption{AI-assisted scientific software workflow and requirements for sustainable scientific stewardship. AI can assist at multiple stages of an iterative workflow, while understanding and scientific judgment, validation and quality, provenance and reproducibility, maintainability and reuse, resource responsibility, and sustainable access and tool independence remain cross-cutting requirements.}\label{fig:workflow}
\end{figure}

Figure~\ref{fig:workflow} summarizes this perspective across the scientific-software lifecycle, where AI can contribute at multiple stages.

\section{What should change in software training?}\label{sec:4}

Direct empirical evidence on AI-aware HEP software training remains limited. Existing HEP studies are largely descriptive or self-reported, but consistently emphasize structured hands-on work, reusable materials, expert interaction, and active learning \citep{r1,r2,r16}. In a recent particle-physics course that encouraged AI on research-shaped assignments, students remained engaged and produced ambitious work, yet foundational learning was not reliably secured \citep{r17}. We therefore combine this emerging HEP evidence with stronger results from AI-assisted programming and computing education.

AI-aware software training has two complementary components: learning \textbf{about AI} and learning \textbf{with AI}. Dedicated instruction should cover assistant and agent capabilities, differences among services, data and credential handling, and the implications of systems that can execute tools rather than only suggest code. But this is not sufficient. AI should also be integrated into established software training, so that familiar competencies such as testing, debugging, version control, reproducibility, profiling, documentation, and resource management are practiced in AI-assisted workflows. Training should also teach researchers to calibrate scrutiny to consequence: code affecting authentication or authorization, data integrity, physics selections, or numerical results warrants stronger independent review than lower-consequence presentation code. The aim is not a separate AI curriculum, but to make AI-assisted work one context in which durable software competencies are taught.

Table~\ref{tab:training} illustrates this integration. An assistant can implement a scientific feature and propose tests while participants design independent validation; participants can also compare AI-generated implementations for numerical equivalence, performance, and resource use. In both cases, AI creates situations in which participants practice durable competencies rather than merely learning to operate the tool. The table translates Figure~\ref{fig:workflow}’s requirements into competencies and adds security and permissions as an agent-specific operational competency.

\begingroup
\small\raggedright\setlength{\tabcolsep}{5pt}\renewcommand{\arraystretch}{1.12}
\setlength{\LTleft}{0pt}\setlength{\LTright}{0pt plus 1fill}\setlength{\LTcapwidth}{\textwidth}
\makeatletter
\long\def\LT@makecaption#1#2#3{%
  \LT@mcol\LT@cols c{\hbox to\z@{\hss\parbox[t]\LTcapwidth{%
    \normalsize\textbf{#2.}\enskip #3\par\vskip\baselineskip}%
  \hss}}}
\makeatother
\begin{longtable}{@{}>{\raggedright\arraybackslash}p{0.18\textwidth}>{\raggedright\arraybackslash}p{0.355\textwidth}>{\raggedright\arraybackslash}p{\dimexpr0.465\textwidth-20pt\relax}@{}}
\caption{Durable scientific-software competencies, ways in which AI changes the challenge of exercising them, and example activities for AI-aware training. The activities are proposed training designs informed by recent evidence on AI-assisted programming and by experience with hands-on sustainable-computing training; they are not presented as experimentally established pedagogical best practices.}\label{tab:training}\\
\toprule
\textbf{Durable competency} & \textbf{Why AI changes the challenge} & \textbf{Example training activity}\\\midrule
\endfirsthead
\multicolumn{3}{@{}l}{\textbf{Table 1 (continued)}}\\[4pt]
\toprule
\textbf{Durable competency} & \textbf{Why AI changes the challenge} & \textbf{Example training activity}\\\midrule
\endhead
\midrule\endfoot
\bottomrule\endlastfoot
\textbf{Understanding and scientific judgment} & Successful code generation or task completion does not ensure that the researcher can explain consequential implementation choices, assumptions, or the evidence supporting the result. & Ask participants to explain an AI-assisted solution, identify assumptions and failure modes, and modify a consequential part of it without simply regenerating the answer.\\
\addlinespace[5pt]
\textbf{Validation and quality} & Generated code can be plausible and executable while remaining incomplete, scientifically inappropriate, or wrong in untested cases; AI-generated tests or validation can fail for the same reason. & Have an assistant implement a scientific feature and propose tests; participants must identify missing cases and construct at least one independently motivated validation, such as comparison with a reference implementation, an analytic limit, an exact-null closure test, or a known invariant.\\
\addlinespace[5pt]
\textbf{Provenance and reproducibility} & Assistants and agents may make many intermediate changes or invoke external tools, making the final code alone an incomplete record of the workflow. & Require participants to preserve sufficient information to reconstruct an AI-assisted workflow, including consequential inputs, actions, dependencies, and validation steps.\\
\addlinespace[5pt]
\textbf{Maintainability and reuse} & Low-cost generation can encourage unnecessary features, duplicated implementations, or repeated computation, increasing complexity faster than the capacity to understand and maintain it. & Ask participants first whether a new implementation or computation is needed, identify existing software or reusable data products where appropriate, and review generated code for documentation, interfaces, ownership, reuse, and future maintenance.\\
\addlinespace[5pt]
\textbf{Resource responsibility} & AI can generate working implementations without judging whether they scale sensibly; faster code is not necessarily more resource- or energy-efficient overall. & Compare alternative implementations for runtime and resource use, test first at small scale, and estimate whether an optimization is worthwhile for the expected workload.\\
\addlinespace[5pt]
\textbf{Sustainable access and tool independence} & Model capabilities, subscription tiers, application programming interfaces, and institutional infrastructure vary, while specific products can change rapidly. & Design core exercises to work with multiple supported tools and assess transferable skills rather than proficiency with one commercial interface.\\
\addlinespace[5pt]
\textbf{Security and permissions} & Agents can read files, execute commands, and interact with services rather than merely suggest code, increasing the consequences of excessive privileges or untrusted inputs. & Give an agent only the resources required for a task and require explicit approval for consequential actions; discuss which permissions are unnecessary.\\
\end{longtable}
\endgroup
\FloatBarrier

The DESY/UHH workshop illustrates how dedicated AI instruction can complement established software training. Its 2026 AI session covered institutional AI services and coding agents, credential handling, permissions, data leakage, and safe use on local and shared infrastructure, connecting AI-specific risks to testing, reproducibility, and responsible resource use. Open-ended feedback included requests for focused training on AI-assisted coding, including safe setup and responsible use.

CMSDAS provides a collaboration-specific setting for integrated practice in authentic analysis tasks. The January 2026 school demonstrated agentic AI for HEP event simulation \citep{r18}. A future CMSDAS physics-analysis exercise could make stewardship explicit: an assistant might modify event selections, write plotting code, or diagnose failures, while participants submit an independent comparison with a trusted baseline or closure test, a short provenance record of consequential AI-assisted actions and validation, and justification of one consequential choice. A compact rubric could assess scientific correctness, independent validation, sufficient provenance, and explanation of assumptions and failure modes. This preserves the authentic task while shifting evidence of competence from code production toward scientific command and validation.

Evidence from programming education increasingly indicates that outcomes depend on how AI assistance is structured. A 72-study review identifies scaffolding and its planned withdrawal as promising responses to over-reliance \citep{r8}, while a randomized seven-week study of 124 students found better higher-order-thinking and programming-logic outcomes with structured than unguided generative AI use \citep{r19}. In practice, scaffolding can require learners to predict expected behavior before consulting an assistant, explain consequential choices, or diagnose deliberately introduced errors. Assistance can be withdrawn where independent competence is itself the learning objective. These constraints do not recreate an artificially “pre-AI” environment; they clarify which capabilities researchers must retain even with automation. This concern is particularly relevant for early-career researchers, for whom persistent delegation could leave foundational skills underdeveloped \citep{r17}.

The evidence used to judge competence should change accordingly. In formal assessment, exercise design, or facilitator feedback, a correct plot, successful compilation, or completed workflow is weak evidence on its own that the learner can justify the computation. More informative evidence includes a validation strategy, explanation of assumptions and failure modes, sufficient provenance, and justification of computational choices. Such assessment is also more demanding for trainers than checking task completion. At schools with tens of participants, structured rubrics, group review, and automated checks can cover routine requirements while reserving expert attention for consequential scientific judgments; facilitators themselves need enough AI literacy to recognize relevant failure modes. Because models and interfaces change rapidly, exercises should emphasize transferable practices and, where feasible, work across multiple supported tools.

The central change is therefore not simply to add more AI content. It is to redesign selected software-training activities so that researchers practice taking scientific responsibility for work produced partly with AI: establishing when an output is correct, understanding consequential choices, preserving sufficient provenance, controlling unnecessary complexity, and deciding whether the resulting workflow is reproducible, maintainable, and computationally sensible.

\section{Where should training happen? Complementary institutional roles}\label{sec:5}

The training model spans general scientific-software competencies and experiment-specific judgment, so no single setting suits all objectives. Institutions can provide tool-agnostic foundations, supported AI services, shared infrastructure, and local guidance on data handling, authentication, policy, and resource use. Collaborations can place these competencies in realistic analysis workflows and establish conventions for provenance, permissions, review, and approval. Furthermore, community efforts can reduce duplication through reusable materials and common practices. The HEP Software Foundation and the Institute for Research and Innovation in Software for High Energy Physics (IRIS-HEP) already provide open training spanning foundational software engineering and HEP-specific tools \citep{r16}, offering a natural setting for piloting and maintaining AI-aware modules. At the September 2026 Community Laboratory for AI-native Research In Physics (CLARIPHY) collaboration meeting, the Education, Training, and Workforce Development breakout identified level-appropriate common curricula, validation and cross-checks, shared introductory resources, and coordination across existing training initiatives as priorities \citep{r22}. These roles need not be rigidly separated; the goal is to connect transferable competencies with the domain-specific settings where researchers exercise scientific judgment.

\section{Discussion and conclusion}\label{sec:6}

AI does not eliminate the need for established scientific-software competencies; it changes the circumstances in which they must be exercised. When researchers can obtain working code or computational results without constructing every part themselves, the existence of a successful output becomes weaker evidence that they command the consequential assumptions, choices, and validation underlying it. Training must therefore place greater weight on the ability to validate, explain, reproduce, maintain, and justify AI-assisted scientific work.

For HEP, the most useful response is not a separate AI curriculum layered on top of existing software training. Researchers need dedicated instruction on the capabilities and constraints of the systems available to them, while AI should also be integrated into authentic software and analysis tasks. Testing, debugging, provenance, resource judgment, and scientific validation can all be taught in this setting, with competence judged increasingly by the evidence supporting a result rather than by code production alone.

This shift toward scientific stewardship is also central to making AI-assisted scientific software sustainable. Sustainability requires more than making today’s task easier or faster: the resulting workflows must remain understandable, reproducible, maintainable, and computationally responsible, while the competencies needed to use AI effectively should not depend unnecessarily on privileged access to particular tools or infrastructure. The durable objective is therefore not proficiency with a particular model or interface, but the ability to take responsibility for computational work as those tools change. Embedding that objective across institutional, collaboration-specific, and community training can help HEP adopt AI while preserving the scientific and computational practices on which sustainable research depends.

\section*{Data availability statement}
No research datasets were generated or analyzed for this study. The workshop feedback discussed in this article was collected for internal workshop evaluation and is not publicly available to protect participant privacy.

\section*{Author contributions}
JA: Conceptualization, Investigation, Visualization, Writing -- original draft, Writing -- review \& editing.

\section*{Funding}
The author acknowledges support from DESY (Hamburg, Germany), a member of the Helmholtz Association HGF, and support by the Deutsche Forschungsgemeinschaft (DFG, German Research Foundation) under Germany’s Excellence Strategy -- EXC 2121 "Quantum Universe" -- 390833306.

\section*{Conflict of interest}
The author declares that the research was conducted in the absence of any commercial or financial relationships that could be construed as a potential conflict of interest.

\section*{Acknowledgments}
During preparation of this manuscript, the author used OpenAI ChatGPT (GPT-5.6 Sol; OpenAI; ChatGPT web application accessed in 2026 at chatgpt.com) to support literature discovery and verification, critical review, drafting, and editing. ChatGPT Work (GPT-6 Astra; OpenAI; accessed in 2026 at chatgpt.com) was used for technical manuscript assembly and preparation of the original vector figure. All AI-assisted text, citations, references, and figure content were reviewed and verified by the author, who takes full responsibility for the final manuscript.

\renewcommand{\bibfont}{\small}
\setlength{\bibsep}{3pt}

\end{document}